\documentclass{svmult}

\usepackage{mathptmx}       
\usepackage{helvet}         
\usepackage{courier}        
\usepackage{type1cm}        
\usepackage{makeidx}         
\usepackage{graphicx}        
\usepackage{multicol}        
\usepackage[bottom]{footmisc}

\makeindex             
\begin{document}

\title{The initial conditions of star formation: cosmic rays as the fundamental regulators}
\titlerunning{The initial conditions of star formation: cosmic rays as the fundamental regulators}
\author{Padelis P. Papadopoulos and Wing-Fai Thi}
\authorrunning{The initial conditions of star formation: cosmic rays as the fundamental regulators} 
\institute{Padelis P. Papadopoulos \at Max Planck Institute for
Radioastronomy, Auf dem Huegel 69, D-5312, Bonn,  Germany, \\ \email{padelis@mpifr-bonn.mpg.de}
\and
Wing Fai-Thi \at UJF-Grenoble 1 / CNRS-INSU, Institut de Planétologie et d'Astrophysique (IPAG)
 UMR 5274, 38041, Grenoble, France, \\ \email{wing-fai.thi@obs.ujf-grenoble.fr}}
%
%
\maketitle


\abstract{Cosmic  rays  (CRs)  control  the  thermal,  ionization  and
  chemical state of the dense  H$_2$ gas regions that otherwise remain
  shielded  from  far-UV  and  optical stellar  radiation  propagating
  through  the dusty  ISM of  galaxies.   It is  in such  CR-dominated
  regions (CRDRs) rather than Photon-dominated regions (PDRs) of H$_2$
  clouds where  the star formation initial conditions  are set, making
  CRs the ultimate star-formation feedback factor in galaxies, able to
  operate  even  in their  most  deeply dust-enshrouded  environments.
  CR-controlled  star formation initial  conditions naturally  set the
  stage for  a near-invariant stellar  Initial Mass Function  (IMF) in
  galaxies as  long as  their average CR  energy density  $\rm U_{CR}$
  permeating their molecular ISM remains  within a factor of $\sim $10
  of its Galactic value.  Nevertheless, in the extreme environments of
  the  compact  starbursts  found  in  merging  galaxies,  where  $\rm
  U_{CR}$$\sim   $(few)$\times    $$10^{3}$\,$\rm   U_{CR,Gal}$,   CRs
  dramatically alter the initial conditions of star formation.  In the
  resulting extreme  CRDRs H$_2$ cloud fragmentation  will produce far
  fewer low  mass ($<$8\,$\rm M_{\odot}$) stars,  yielding a top-heavy
  stellar  IMF.   This will  be  a  generic  feature of  CR-controlled
  star-formation  initial conditions,  lending a  physical base  for a
  bimodal  IMF  during galaxy  formation,  with  a  top-heavy one  for
  compact merger-induced starbursts, and an ordinary IMF preserved for
  star  formation in  isolated  gas-rich disks.   In  this scheme  the
  integrated galactic IMFs (IGIMF) are expected to be strong functions
  of  the star  formation  history of  galaxies.   Finally the  large,
  CR-induced,  ionization  fractions  expected  for  (far-UV)-shielded
  H$_2$  gas in  the CRDRs  of  compact starbursts  will lengthen  the
  ambipolar  diffusion  (AD)  timescales  so  much as  to  render  the
  alternative   AD-regulated   rather   (Jeans  mass)-driven   star
  formation  scenario  as utterly  unrealistic  for  the  ISM in  such
  galaxies.}

\section{Cosmic rays versus far-UV and optical photons}
\label{sec:1}

\begin{figure}
\sidecaption
\includegraphics[scale=0.41]{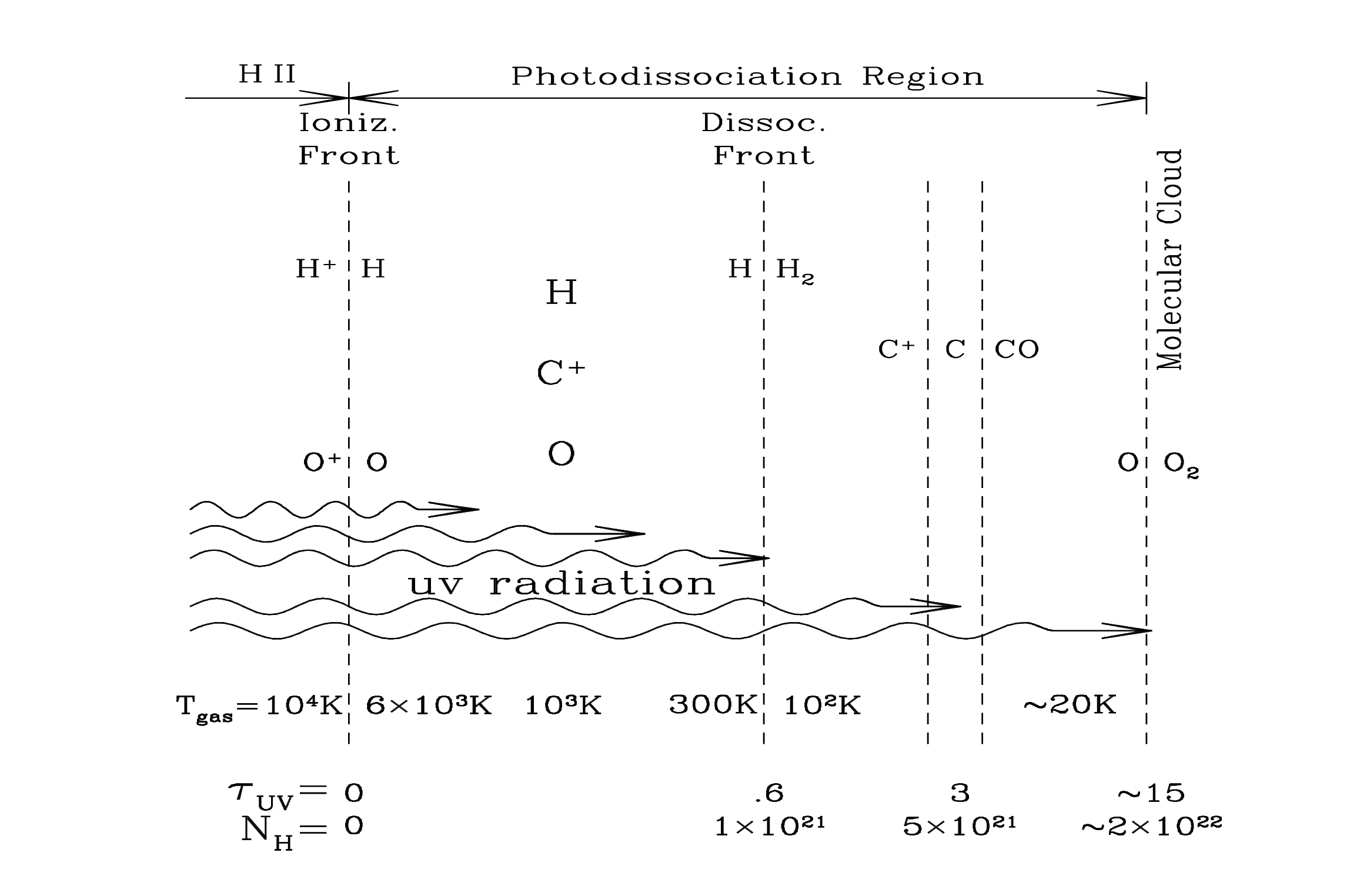}
%
%
\caption{The  structure  of a  typical  PDR/CRDR.  Atomic and  ionized
  species reside on the  (far-UV)-illuminated outer parts (left) while
  neutral atoms (O)  and molecules such as CO  dominate deeper inside,
  the domain of CRDRs and  CR-controled thermal and chemical gas
  states.}
\label{fig:1}       
\end{figure}

Much  of the  interstellar medium  (ISM)  in galaxies  resides in  the
so-called Photon-dominated regions (PDRs) where stellar far-UV photons
(6\,eV$<$$h\nu  $$<$13.6\,eV) determine  the  thermal, ionization  and
chemical  state of  atomic  (HI) and  molecular  (H$_2$) hydrogen  gas
\cite{HolTiel99}.   For HI these  phases are  the Warm  Neutral Medium
(WNM,    with    $\rm     T_{kin}$$\sim    $$10^4$\,K    and    n$\sim
$(0.1-1)\,cm$^{-3}$)  and  the Cold  Neutral  Medium  (CNM, with  $\rm
T_{kin}$$\sim    $(80-200)\,K    and   n$\sim    $(10-100)\,cm$^{-3}$)
\cite{Wolf03}.  For  the molecular PDRs  in H$_2$ clouds on  the other
hand:   $\rm    n$$\sim   $(10$^{2}$-10$^{6}$)\,cm$^{-3}$   and   $\rm
T_{kin}$$\sim  $(30-100)\,K.   Such PDRs  start  from  the surface  of
molecular clouds (MCs)  with a layered distribution (Fig.~\ref{fig:1})
of atomic and ionized species on  the outside (HI, C$^{+}$, C, O), and
continue   deep  inwards   where   molecules  such   as  CO   dominate
\cite{TiHol95a,  TiHol95b}, marking  the location  of such  clouds via
their mm/submm rotational line emission.  The thermal, ionization, and
chemical state of all these phases is determined by the far-UV photons
whose influence, especially on the ISM chemistry, can extend very deep
inside the low-density  MCs ($\rm n(H_2)$$\sim $(100--500)\,cm$^{-3}$)
typical in the  Galaxy.  The dust in PDRs is  heated via absorption of
the  ambient far-UV  and  optical radiation  and  cools via  continuum
thermal  IR/submm  emission while  the  gas  is  heated via  electrons
ejected by the photolectric effect  on the concomitant dust grains and
polycyclic  aromatic hydrocarbons (PAHs).   Despite the  gas receiving
$\sim $10$^{2}$--10$^{3}$ \emph{less} heating per unit volume than the
dust, in  PDRs it is  always $\rm T_{kin}$$>$$\rm T_{dust}$.   This is
because of  the much  less efficient cooling  of the gas  via spectral
lines while dust cools via  continuum radiation across a wide range of
wavelengths  (typically peaking  at IR/submm  wavelengths).  Turbulent
heating  of   the  gas  in   MCs  will  further  reinforce   the  $\rm
T_{kin}$$>$$\rm T_{dust}$  inequality, since  it does not  involve the
dust reservoir.

For  MCs in  the typical  ISM conditions  of spiral  disks  where $\rm
\langle A_v \rangle$$\sim $3-7 (in units of optical extinction), there
can  only   be  very  little   H$_2$  gas  residing  in   their  dense
(far-UV)/optical-shielded regions.  Its mass  fraction per MC is $\sim
$(1-5)\%, a census  found from surveys of CO  J=1--0 (total H$_2$ mass
indicator) and  rotational transitions  of dense gas  tracer molecules
such as  the HCN (mostly J=1--0),  and using the HCN/CO  line ratio to
determine  $\rm  f_{dense}$=$\rm M(n\geq  10^4\,cm^{-3})/M_{tot}(H_2)$
\cite{Sol92,  GaoSol04}.  Here  is  important to  note  that the  more
complex heavy rotor  molecules such as HCN will  necessarily trace the
dense  but also  (mainly)  the (far-UV)-shielded  mass  of MCs  since,
unlike  H$_2$  and  its  tracer  molecule CO,  they  have  much  lower
dissociation potentials  ($\sim $5\,eV  for HCN versus $\sim $13\,eV
for H$_2$, and $\sim $10\,eV for~CO).

\begin{figure}
\sidecaption
\includegraphics[scale=0.45]{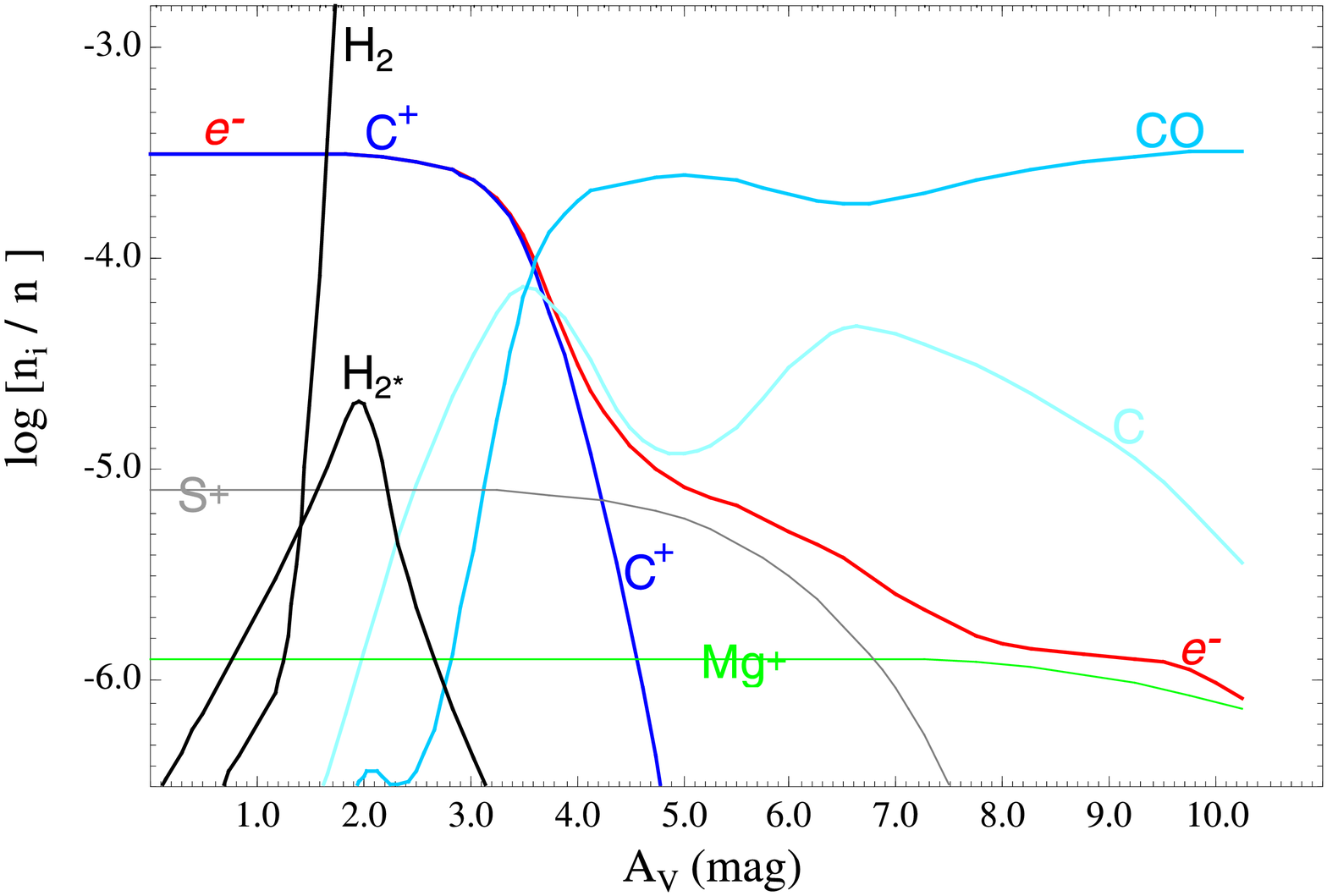}
%
%
\caption{The PDR$\rightarrow  $CRDR transition  zone as marked  by the
  rapidly  decreasing C$^{+}$  and  e$^{-}$ abundances  deep inside  a
  (far-UV)-illuminated molecular cloud  (radiation field incoming from
  the left).  Note the dramatic  fall of the electron abundance beyond
  $\rm A_v$$\sim $4.}
\label{fig:2}       
\end{figure}

Simple  approximate estimates of  the mass  fraction contained  in the
PDRs of (far-UV)-irradiated MCs can  be found by computing the mass of
outer atomic  HI layer  marking the HI$\rightarrow  $H$_2$ transition,
itself  comparable  to  the mass  of  the  warm  PDR H$_2$  gas  layer
extending further  inwards.  Thus the PDR-residing  gas column density
(HI and H$_2$) will be  $\sim $2$\times $$\rm N_{tr}(HI)$.  The latter
can be computed for a given radiation field $\rm G_{\circ}$ (in Habing
units), metallicity Z (Z=1 being solar), average gas density n, and an
H$_2$ formation rate $\rm R_{f}$.  In units of optical extinction this
transition layer is (\cite{Pel06}):

\begin{equation}
\rm A^{(tr)} _{v} = 1.086 \xi^{-1} _{FUV} \ln\left[1+\frac{G_{\circ} k_{\circ}}{n R_f}\Phi\right]
\end{equation}

\noindent
where  $\rm  k_{\circ}$=4$\times  $$10^{-11}$\,s$^{-1}$ is  the  H$_2$
dissociation   rate   (for    $\rm   G_{\circ}$=1),   $\rm   R_f$$\sim
$3$\times  $$10^{-17}$\,cm$^{-3}$\,s$^{-1}$   its  formation  rate  on
grains       (for      typical       CNM      HI),       and      $\rm
\Phi$=6.6$\times $$10^{-6}$$\sqrt{\pi}$$\rm Z^{1/2}$$\rm \xi_{FUV}$ is
the  H$_2$  self-shielding   function  integrated  over  the  HI/H$_2$
transition    layer,    with     $\rm    \xi    _{FUV}$=$\rm    \sigma
_{FUV}/\sigma_{V}$$\sim $2-3  being the  dust cross section  ratio for
far-UV and optical light.  Using existing formalism (\cite{Pel06}) the
PDR-related gas mass fraction per MC then~is:

\begin{equation}
\rm f_{PDR} \sim 2\times \left[1-\left(1-\frac{4A^{(tr)} _{v}}{3\langle A_v \rangle}\right)^{3}\right],
\end{equation}

\noindent
assuming spherical and uniform MCs  that do not cross-shield and hence
each receives  the full (far-UV)/optical  radiation field, assumptions
that make  the computed $\rm  f_{PDR}$ a maximum. The  average optical
extinction $\rm \langle A_v \rangle$ per  MC is of course finite and a
function  of  the ambient  ISM  conditions,  mainly  the gas  pressure
(dominated  by  supersonic non-thermal  rather  than thermal  velocity
fields). Expressing an  empirical Galactic (average MC density)-(size)
relation:   n$\propto   $(2R)$^{-1}$,    along   with   its   expected
normalization  in  terms  of   cloud  boundary  pressure  $\rm  P_{e}$
(\cite{Pel06}) in terms of average optical extinction yields:

\begin{equation}
\rm \langle A_v \rangle \sim 0.22\,Z\,\frac{n_{\circ}}{100\,cm^{-3}}
\left(\frac{P_e/k_B}{10^4 cm^{-3} K}\right)^{1/2},
\end{equation}

\noindent
where $\rm n_{\circ}$$\sim $1500\,cm$^{-3}$.  For average MC densities
of $\sim $100\,cm$^{-3}$, Solar  metalicities, $\rm \xi _{FUV}$=2, and
$\rm  G_{\circ}$=5-10  (for  typical  star-forming gas)  Equation  1.1
yields  $\rm   A^{(tr)}  _{v}$$\sim  $0.50-0.75,   while  for  typical
interstellar   pressures   in   galactic  disks   $\rm   P_e/k_B$$\sim
$(1-2)$\times  $10$^{4}$\,K\,cm$^{-3}$   it  would  be   $\rm  \langle
A_v\rangle $$\sim  $3.3-4.7. The  latter is the  $\rm A_v$  range over
which the abundance of C$^{+}$ plummets, giving rise to C and CO while
the free electron abundance is much reduced (see Fig.~\ref{fig:2}).

  For  the finite  MCs  this  ´outer´ HI/(warm  H$_2$)  PDR gas  still
  contains $\rm  f_{PDR}$$\sim $0.74--1  of their mass  (from Equation
  1.2),  i.e.   \emph{most  of  the   HI  and  H$_2$  gas  in  typical
    star-forming ISM environments resides  in PDRs.} Later we will see
  that this will not be so  in the high-pressure ISM of Ultra Luminous
  Infrared  galaxies (ULIRGs),  merger systems  with  very IR-luminous
  ($\rm   L_{IR}$$\geq  $10$^{12}$\,L$_{\odot}$)   compact  starbursts
  (Sect.~\ref{sec:3}).

\section{The  physical conditions of the gas: PDRs versus CRDRs}
\label{sec:2}

The wide range of physical conditions  of the HI and H$_2$ gas in PDRs
(Sect.~\ref{sec:1})  is the direct  result of:  a) the  widely varying
strengths of the dust-absorbed  far-UV and optical radiation (and thus
of the gas  photoelectric heating per H, and  dust heating) within the
PDRs, and  b) the large  range of densities  (and thus of  the cooling
power per  H) that  are expected in  the supersonically  turbulent MCs
(\cite{PaNo02}).  Dissipation of  the supersonic turbulence via shocks
(\cite{PanPad09})  will  further  contribute  to  the  wide  range  of
temperatures observed  in the PDR-residing gas of  MCs. Moreover their
average electron  abundance $\rm x(e)$=$\rm  n_e/[2n(H_2)]$ changes by
$\sim  $2-3 orders  of magnitude,  starting  from a  maximum of  $\sim
$3$\times $$10^{-4}$ (that corresponds to a fully ionized~carbon).

 The dense  gas cores ($\rm  n$$\sim $(10$^{5}$--10$^{6}$)\,cm$^{-3}$)
 in Galactic  CRDRs on the  other hand span  a much narrower  range of
 temperatures    and   ionization    states,    with   $\rm    \langle
 T_{kin}\rangle $$\sim  $10\,K (and $\sim $30\%  dispersion), and $\rm
 x(e)$$\sim $(few)$\times $$10^{-8}$ (varying by less than a factor of
 $\sim  $5).  These  CRDR conditions,  widely observed  in  the Galaxy
 (\cite{BerTaf07}, \cite{Pin07}),  are a  result of a  nearly complete
 lack  of  far-UV  and  optical  photons  in  such  regions  (thus  no
 photolectric  gas heating  or strong  dust heating),  nearly complete
 dissipation of  supersonic gas motions (\cite{Pin10}),  and the onset
 of strong gas-dust thermal  coupling setting $\rm T_{kin}$$\sim $$\rm
 T_{dust}$.   This simply means  that cold  dust grains  inside CRDRs,
 warmed only by the feeble IR  radiation able to leak inside them, can
 now act  as powerful 'thermostats' of  the H$_2$ gas.   The latter is
 now heated  almost exclusively  by primary and  secondary ionizations
 induced by low-energy CRs (10\,MeV$\leq $$\rm E_{CR}$$\leq $100\,MeV)
 \cite{Gold78}  which, unlike  far-UV and  optical  photons, penetrate
 deep  inside  such  regions,  and  thus  \emph{fully  regulate  their
   thermal, ionization and chemical states.}

The aforementioned  CR-controlled dense  H$_2$ gas states  in ordinary
star-forming  environments  is  expected  also  on  broad  theoretical
grounds, namely: a) high density  gas regions will be typically nested
well inside  less dense gas structures of  supersonic MCs \cite{Oss02}, 
and  thus  will  always  'see' a  much-attenuated  stellar  radiation
field,  b)  high-density gas  is  also  strongly  cooling gas  (since
(line-cooling)$\propto$$\rm  [n(H_2)]^{2}$), and  c)  turbulent energy
injected  on the  largest gas  structures with  low  average densities
$\sim  $(10$^{²}$-10$^{3}$)\,cm$^{-3}$ will  have fully  dissipated in
the high  density gas  of CRDRs. It  is actually these  three reasons,
along  with the  strong thermal  coupling  of the  gas in  CRDRs to  a
necessarily cold dust (for  lack of far-UV/optical photons) reservoir,
that \emph{make it  very difficult for large masses  of warm and dense
  H$_2$ gas to exist  in ordinary star-forming galaxies.}  Finally, as
already mentioned  in \ref{sec:1}, only  few\% of the total  H$_2$ gas
mass resides at  such high densities anyway, as  expected also from MC
simulations  (\cite{PaNo02})   under  conditions  found   in  ordinary
star-forming spiral disks.

 \subsection{Computing $ T_{kin}$ and x(e) in CRDRs: a simpler method }

 In  CRDRs the  $\rm T^{(min)}  _{kin}$ and  $\rm  x_{min}(e)$ minimum
 values  expected  for  H$_2$   gas  in  galaxies  are  set.   However
 determining them  involves solving the coupled  equations of thermal,
 chemical and  ionization balance in  such regions since CRs  heat the
 gas but  also affect  the abundance of  various molecular  and atomic
 coolants.  We can nevertheless make some simplifying assumptions that
 will keep  the physics  involved transparent while  demonstrating the
 range  of  $\rm T^{(min)}  _{kin}$  and  $\rm x_{min}(e)$.  Following
 standard methods (\cite{Gold01}) $\rm  T^{(min)} _{kin}$ in CRDRs can
 be computed by solving the equation of thermal balance:

\begin{equation}
\rm \Gamma _{CR} = \Lambda _{line}+\Lambda _{g-d},
\end{equation}
  
\noindent
where $\rm  \Gamma _{CR}$$\propto $$\rm  \zeta_{CR} n(H_2)$ is  the CR
heating  with  $\rm \zeta  _{CR}$$\propto  \rm  U_{CR}$  being the  CR
ionization rate, and $\rm U_{CR}$  the average CR energy density.  For
$\rm  T_{kin}$$\leq $50\,K (which  encompasses what  is expected  for
CRDRs of  ordinary star-forming galaxies),  the line cooling  term $\rm
\Lambda _{line}$ is dominated by  CO lines.  However for the much more
extreme CRDRs expected in ULIRGs, higher temperatures are possible and
thus cooling  from the atomic  fine structure lines O\,I  (at 63\,$\mu
$m) and C$^{+}$  (at 158\,$\mu $m) can become  important.  The C$^{+}$
abundance in  CRDRs is not necessarily  negligible as it  is no longer
controlled by  a stellar far-UV radiation  field (as in  PDRs), but by
ISM-CR interactions  inducing internal UV radiation  which destroys CO
and produces C$^{+}$, a clear demonstration of the coupled thermal and
chemical states  of the  gas in CRDRs.  Finally the term  $\rm \Lambda
_{g-d}$ denotes  the all-important (in  CRDRs) cooling of the  gas via
the   gas-dust  coupling   which   depends  on   the   gas  and   dust
temperatures. Thus the thermal balance equation becomes

\begin{eqnarray}
\rm \Lambda _{CO}(T_{kin}) + \Lambda _{g-d}(T_{kin},T_{dust})+ 
 \Lambda _{OI\,63}(T_{kin}) + \Lambda _{C^{+}}(T_{kin})=\Gamma _{CR},
\end{eqnarray}

\noindent
(see  the Appendix  for the  detailed expressions  used). We  have not
considered turbulent gas heating since  we want to compute the minimum
CR-controlled  gas temperatures and  since supersonic  turbulence (and
thus shock-heating) has typically  dissipated in the dense pre-stellar
gas regions inside CRDRs.

\begin{figure}
\sidecaption
\includegraphics[scale=0.50, angle=90]{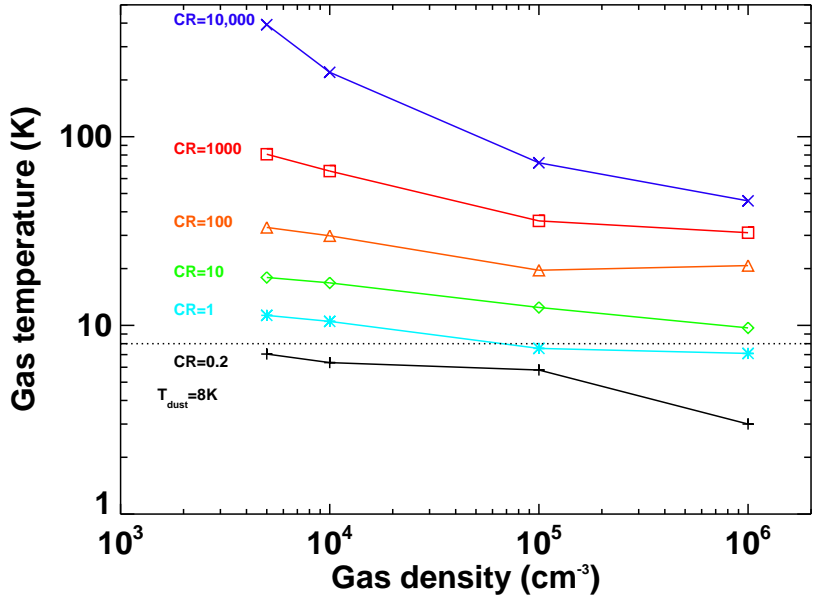}
\caption{The equilibrium temperature  of (far-UV)-shielded regions in
  CRDRs computed from equation 1.5. The CR energy densities range from
  those  expected in ordinary  star-forming environments  Galaxy [$\rm
    U_{CR}$=(0.2-10)$\times  $$\rm U_{CR,Gal}$]  to those  expected in
  the        extreme        CRDRs        of        ULIRGs        [$\rm
    U_{CR}$=($10^2$--$10^4$)$\times   $$\rm   U_{CR,Gal}$].   A   dust
  temperature  of $\rm  T_{dust}$=8\,K, and  a Galactic  CR ionization
  rate of  $\rm \zeta _{CR}$=5$\times  $10$^{-17}$\,s$^{-1}$ have been
  assumed.}
\label{fig:3}       
\end{figure}

In order  to solve  the gas thermal  balance equation in  principle we
also  must solve for  $\rm T_{dust}$  in CRDRs.   For the  purposes of
computing $\rm T^{(min)} _{kin}$  we set $\rm T_{dust}$=8\,K, which is
typical  for  CR-dominated  cores  in  the Galaxy.   Any  stronger  IR
radiation field leaking inside  such cores (as expected in starbursts)
or remnant turbulent  gas heating can only raise  the gas temperatures
computed here.  In Fig.~\ref{fig:3}  the $\rm T^{(min)} _{kin}$ values
show  that CR-heated dense  gas ($>$10$^{4}$\,cm$^{-3}$)  remains cold
($\sim     $(10-15)\,K)    for     $\rm    U_{CR}$     within    $\sim
$(0.2-10)$\times $$\rm  U_{CR,Gal}$, rising decisively  only when $\rm
U_{CR}$$\sim  $(few)$\times   $$10^{3}$\,$\rm  U_{CR,Gal}$  where  the
thermal   balance   equation   yields  $\rm   T^{(min)}   _{kin}$$\sim
$(40-100)\,K.

\begin{figure}
\sidecaption
\includegraphics[scale=0.50, angle=90]{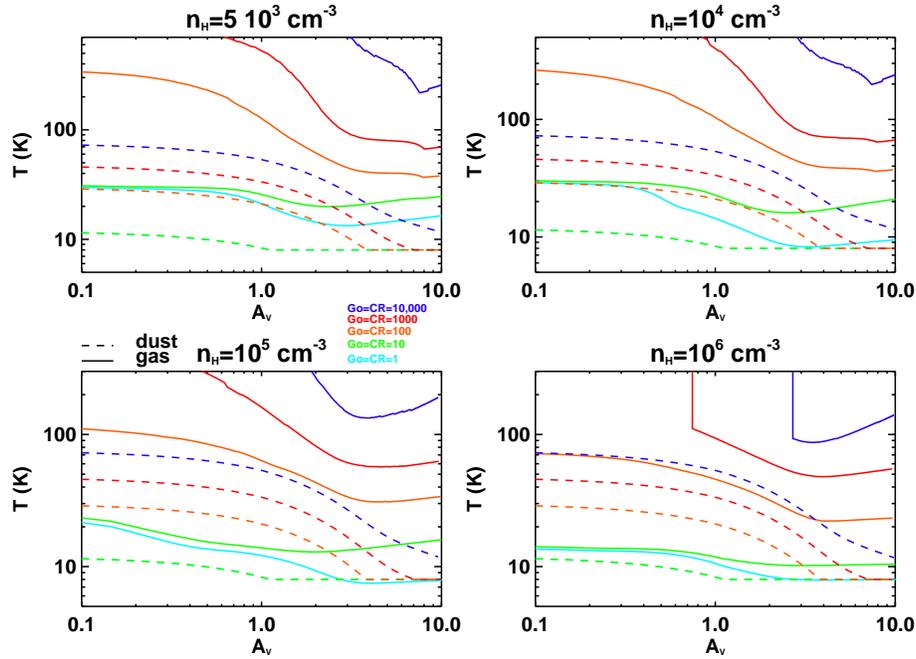}
\caption{The gas and dust temperature profiles (solid and dotted lines
  of  similar  color) versus  optical  extinction  inside a  molecular
  cloud.   These were  computed  by solving  the  coupled thermal  and
  chemical  balance equations (\cite{Pap11}),  while also  assuming a
  far-UV  radiation field (incident  on the  $\rm A_{v}$=0  surface of
  plane clouds) that scales in  an identical fashion as the average CR
  energy density pervading the  H$_2$ cloud. Their common scaling runs
  as    (1,10,$10^2$,$10^3$,$10^4$)$\times   $Galactic.    Unlike   in
  Fig.~\ref{fig:3} the  dust is  significantly warmed by  the stronger
  (far-UV)/optical radiation fields absorbed by the outer cloud layers
  and the re-radiated dust IR emission that penetrates deeper into the
  cloud.}
\label{fig:4}       
\end{figure}

Solving for  the coupled chemical and  thermal states of  dense gas in
CRDRs  does not  change much  of the  aforementioned picture  while it
makes  clear another  important aspect  that of  CR-controlled thermal
states of  the gas,  namely that high  CR energy  densities \emph{will
  induce a  strong gas-dust  thermal decoupling}, with  $\rm T^{(min)}
_{kin}$ remaining  significantly higher  than $\rm T_{dust}$,  even at
high gas densities (Fig.~\ref{fig:4}).

This happens  even as the  dust temperatures deep inside  H$_2$ clouds
rise  because  of an  incident  radiation that  is  now  set to  scale
identically   to  the   average  CR   energy  density   (both  assumed
proportional  to  the average  star-formation  rate density  $\dot\rho
_{sfr}$).  This CR-induced thermal decoupling persists deep inside the
cloud  ($\rm A_v$=5--10) even  at the  highest gas  densities explored
here where gas-dust thermal interaction will be the strongest. It must
also be noted that for a  given average $\dot\rho _{sfr}$ in a galaxy,
the  resulting gas-dust  temperature  inequality will  likely be  even
larger since cross-shielding between H$_2$ clouds as well as dusty WNM
and  CNM HI  gas interdispersed  between such  clouds will  reduce the
average  radiation  field  incident   on  them,  but  not  the  deeply
penetrating CRs.  Thus  while $\rm U_{CR}$$\propto $$\dot\rho _{sfr}$,
the scaling of average $\rm  G_{\circ}$ with $\dot\rho _{sfr}$ will be
weaker than  linear (unlike what  was assumed in  Fig.~\ref{fig:4} for
simplicity).

 From Fig.~\ref{fig:4} it can also be discerned that even for strongly
 boosted  far-UV   radiation  fields,  gas   temperatures  are  solely
 CR-controlled and remain almost  invariant with $\rm A_v$ beyond $\rm
 A_V$$\sim $5.   This characteristic of  CRDRs is of  great importance
 when it  comes to the initial  conditions of star  formation, and the
 near-invariant mass scale of the  stellar IMF found in the Galaxy and
 ordinary star-forming~galaxies.

 An  order  of magnitude  estimate  of  the CR-regulated  equilibrium
 ionization  fraction  of  (far-UV)-shielded  gas can  also  be  found
 without resorting to solving the full chemistry network involved. 
 Following an analytical approach (\cite{McKee89}) yields

\begin{eqnarray}
\rm x(e) = 2\times 10^{-7} r^{-1} _{gd} \left(\frac{n_{ch}}{2n(H_2)}\right)^{1/2}
\left[\left(1+\frac{n_{ch}}{8n(H_2)}\right)^{1/2} + \left(\frac{n_{ch}}{8n(H_2)}
\right)^{1/2}\right]
\end{eqnarray}

\noindent
\noindent
where   $\rm   n_{ch}$$\sim   $500\,$\rm   \left(r^2   _{gd}\,   \zeta
_{-17}\right)\, cm^{-3}$ is a characteristic density encapsulating the
effects  of  cosmic ray  ionization  and  ambient  metallicity on  the
ionization balance ($\rm r_{gd}$ is the normalized gas/dust ratio $\rm
r_{gd}$=[(G/D)/100]   with   G/D(gas-to-dust   mass)=100   for   Solar
metallicities.   The  main  assumptions  underlying  the  validity  of
Equation 1.6  are three, namely: a) all  the gas is H$_2$  and all the
carbon is locked in CO  (so C$^{+}$ does not dominate the ionization),
b) all negative charge is  carried by the free electrons and molecular
ions are destroyed mostly  by recombination with these free electrons,
and c) the metal abundance is small with respect to the heavy molecule
abundance.

For the Galaxy ($\rm \zeta_{CR,Gal}$=5$\times$10$^{-17}$\,s$^{-1}$) it
is $\rm n_{ch}$=2.5$\times $10$^3$\,cm$^{-3}$, and for a typical dense
core   of   $\rm   n(H_2)$=10$^{5}$\,cm$^{-3}$   it  would   be   $\rm
x(e)_{Gal}$$\sim $2.4$\times  $10$^{-8}$. A boost  of $\rm \zeta_{CR}$
by a  factor of 10 yields $\rm  x(e)$=8.4$\times $10$^{-8}$, retaining
the  low  ionization  fraction  necessary for  an  ambipolar-diffusion
regulated star-formation to proceed in such cores (\cite{McKee89}).

\section{CRDRs and the star formation initial conditions in galaxies}
\label{sec:2}

 CRDRs  are   the  ISM  regions   where  the  initial   conditions  of
 star-formation are  set.  Thus their thermal  and/or ionization state
 is expected to  play a critical role on the  emergence of the stellar
 IMF and  its mass scale.   This remains the  case for both  main star
 formation  scenaria proposed,  namely:  a) star  formation driven  by
 gravitational collapse of high overdensity peaks in supersonically
 turbulent H$_2$ clouds, and  b) regulated by ambipolar-diffusion (AD)
 in dense cores whose low x(e) allows the magnetic field lines to slip
 away  making them  supercritical  (i.e.  $\rm  M_{core}/M_{\Phi}$$>$1
 where  $\rm  M_{\Phi}$=$\rm 0.12\Phi/G^{1/2}$,  and  $\Phi  $ is  the
 magnetic field  flux threading the cores, \cite{Mous76})  and able to
 collapse towards  star formation.  The only difference  will be which
 CRDR  gas  property:  the  Jeans   mass  or  x(e),  is  important  in
 determining the stellar IMF.

It is worth emphasizing that  star formation initial conditions set in
CRDRs naturally explain the  first mystery regarding the stellar IMF,
namely its near-universality in ordinary ISM environments.  Indeed the
large range  of thermal, turbulent, and ionization  states present for
the  H$_2$ gas  phase  in the  Milky  Way, makes  the  emergence of  a
near-universal stellar IMF an unlikely  outcome even for the Galaxy if
such states were  all used as star formation  initial conditions.  The
gas in  CRDRs on  the other  hand, lying beyond  the reach  of varying
far-UV radiation  fields and turbulence-induced shock  heating, with a
near-isothermal   state  at   $\rm   T_{kin}$$\sim  $$\rm   T_{dust}$,
near-thermal  motions,  and the  lowest  possible  x(e)  values, is  a
natural  choice   of  ´well-protected´  initial   conditions  of  star
formation  that can  then lead  to a  near-invariant stellar  IMF mass
scale.  Similar arguments have  been made in the past (\cite{Lars95}),
but only recently it was demonstrated just how robust the cold thermal
state  of  CRDRs  remains   irrespective  of  the  average  conditions
prevailing  at the  outer  boundaries  of the  MCs  that contain  them
(\cite{Elm08}).  Finally, as  long as  the average  CR  energy density
pervading CRDRs remains  within a factor of $\sim  $10 of its Galactic
value,  neither the  temperature  nor the  ionization fraction  within
CRDRs  changes  substantially.   This  state  of  affairs  is  however
dramatically altered  in the compact,  merger-induced starbursts found
in ULIRGs.

\section{CRDRs in compact starbursts: new initial conditions for star formation}
\label{sec:3}

Some of  the most spectacular star  formation events in  the local and
the  distant Universe  are found  in strong  galaxy mergers.   In such
systems tidal  torques funnel the bulk  of the molecular  gas of their
gas-rich  disk  progenitors into  regions  not  much  larger than  the
Galactic Center (\cite{Sak08}).  There they settle into small (2r$\leq
$(100-300)\,pc) and  very turbulent  H$_2$ gas disks,  fueling extreme
star-forming events  (\cite{DS98, Sak08}).   The H$_2$ gas  in ULIRGs,
with   one-dimensional   Mach   numbers   of   $\rm   M_{ULIRGs}$$\sim
$(5-30)$\times  $$\rm M_{spirals}$,  is expected  to have  larger mass
fractions  at densities  of  $>$10$^{4}$\,cm$^{-3}$ (\cite{PanPad09}),
and this is indeed observed (\cite{GaoSol04, Sol92}).

 However most of  the large dense H$_2$ masses  found in ULIRGs cannot
 reside  in PDRs,  a result  of the  high (turbulent)  pressures ($\rm
 P_{e}/k_B$$\sim $$10^{7}$)\,K\,cm$^{-3}$)  and high gas  densities in
 these galaxies.  Indeed, from  Equation 1.3 (and a Solar metallicity)
 such high  gas pressures yield  $\rm \langle A_v\rangle  $$\sim $100,
 while    for    the   high    \emph{average}    gas   densities    of
 $\geq  $$10^4$\,cm$^{-3}$ typical  for ULIRGs  (\cite{Gre09}),  it is
 $\rm A^{(tr)} _v$$\leq $2 (Eq.  1.1) even for far-UV radiation fields
 as strong as $\rm G_{\circ}$=$10^4$.  Thus any warm gas mass fraction
 residing in  PDRs will  be $\leq $15\%  (Equation 1.2).   In summary,
 while much  larger dense gas  mass fractions are expected  and indeed
 observed in  ULIRGs than  in isolated star-forming  gas-rich spirals,
 \emph{most  of  their  dense  gas  mass will  not  reside  in  PDRs}.
 Nevertheless recent  observational results indicate  large amounts of
 warm  \emph{and}   dense  gas  in  some  ULIRGs,   with  average  gas
 temperatures much larger than  those deduced for the concomitant dust
 (\cite{Pap12}).  Since a massive warm and dense H$_2$ phase cannot be
 in PDRs,  far-UV/optical photons  are inadequate energy  suppliers of
 its observed state in ULIRGs.   Only high CR energy densities (and/or
 strong turbulent  heating) is  capable of volumetrically  heating the
 H$_2$ gas  (unlike far-UV photons  in PDRs) while also  maintaining a
 significant $\rm T_{kin}$$>$$\rm T_{dust}$ inequality.

 CR   energy    densities   of   $\rm    U_{CR}$$\sim   $[(few)$\times
   $10$^{2}$--$10^4$]$\times  $$\rm U_{CR,Gal}$  are  expected in  the
 compact starbursts  found in ULIRGs (\cite{Pap10})  and, as indicated
 by Fig.~\ref{fig:4}, such high  $\rm U_{CR}$ can easily maintain high
 gas temperatures while the concomitant dust remains much cooler. Such
 conditions have  been inferred  in the past  for the  Galactic Center
 (\cite{Ys07})   as    well   as   for   the    center   of   NGC\,253
 (\cite{Brad03,Hail08}).   However   in  the  case   of  ULIRGs  these
 conditions  prevail over  galaxy-size H$_2$  gas reservoirs  that are
 $\sim $2-3  orders of magnitude  more massive.  Thus  temperatures of
 $\rm T_{kin}$$\sim  $(50-100)\,K are  the new \emph{minimum  } values
 possible  for gas  in  ULIRGs, and  represent dramatically  different
 initial conditions for star formation in their ISM environments.

The  thermal  state  of  dense  gas  in CRDRs  is  thus  no  longer  a
near-invariant but strongly altered  by the high CR energy backgrounds
expected  in  ULIRGs.  In  the  so-called  gravoturbulent H$_2$  cloud
fragmentation  scenario  (\cite{Kl04})  this large  temperature  boost
across a wide density range (Fig.~\ref{fig:3}) is expected to surpress
low  mass star  formation.   This is  because  the Jeans  mass of  the
CR-heated gas in CRDRs,

\begin{equation}
\rm    \nonumber   M^{(c)}   _{Jeans}=\rm    \left(\frac{k_B   T_k}{G\mu
  m_{H_2}}\right)^{3/2}               \rho_c               ^{-1/2}=\rm
0.9\left(\frac{T_k}{10K}\right)^{3/2}\left[\frac{n_c(H_2)}{10^4
    cm^{-3}}\right]^{-1/2}\, M_{\odot},
\end{equation}

\noindent
changes  from $\rm M^{(c)}  _{Jeans}$$\sim $0.3\,M$_{\odot}$  for $\rm
n_c(H_2)$=10$^{5}$\,cm$^{-3}$,  and $\rm  T_k$=10\,K (typical  for the
Galaxy) to  $\rm M^{(c)} _{Jeans}$$\sim  $(3--10)\,M$_{\odot}$ for the
same density and $\rm T_k$=(50-110)\,K  expected to be the new minimum
in the extreme CRDRs of ULIRGs.  It is worth pointing out that an $\rm
M^{(c)}   _{Jeans}$$\sim    $0.3\,M$_{\odot}$   nicely   matches   the
characteristics mass  of the stellar  IMF (the so-called  IMF `knee`).
Its  observed near  invariance under  a wide  range of  ISM conditions
(e.g.    radiation  fields)   has   been  demonstrated   theoretically
(\cite{Elm08}),  leaving   CR  energy  densities   \emph{as  the  sole
  star-formation  feedback  mechanism that  can  decisively alter  the
  initial conditions  of star formation in  galaxies.} This mechanism,
driven by the  massive stars (whose strong winds  and explosive deaths
provide the CR acceleration), will operate unhindered in the dusty ISM
environments of extreme merger/starbursts (unlike photons), and it can
reach  where  supersonic  turbulence  cannot,  the  small,  dense  and
(far-UV)-shielded cores deep inside H$_2$ clouds.

\begin{figure}
\sidecaption
\includegraphics[scale=0.50, angle=90]{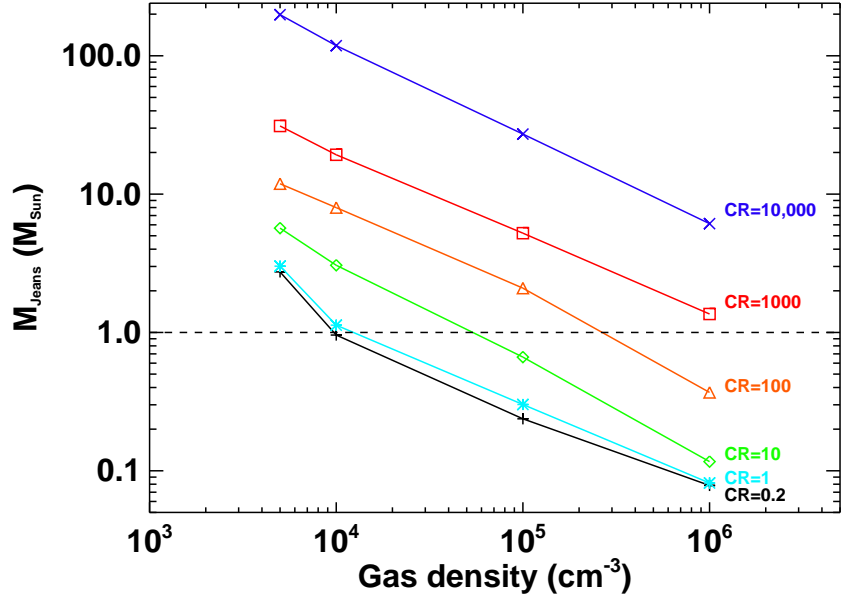}
\caption{The  Jeans  mass estimated  from  Equation  1.7  and the  gas
  temperatures  inside CRDRs,  which  were determined  by solving  the
  fully   coupled  thermal  and   chemical  balance   equations.   The
  temperatures shown are  averages of the $\rm A_v$=5--10  part of the
  H$_2$ clouds where CR become  dominant regulators of the gas thermal
  and chemical states (Fig.~\ref{fig:4}).}
\label{fig:5}       
\end{figure}

In Fig.~\ref{fig:5}  the Jeans  masses are shown  for a  density range
encompassing that of  typical star-forming H$_2$ clouds, demonstrating
that  for  $\rm  U_{CR}/U_{CR,Gal}$$\geq  $$10^{3}$,  $\rm  M_{jeans}$
increases by  a factor of $\sim  $10 across the  entire density range.
This will  invariably lead to higher characteristic  mass $\rm M^{(*)}
_{ch}$  for newly  formed  stars,  and thus  a  top-heavy stellar  IMF
(\cite{Lars95,Elm08}).    For  an   $\rm  M^{(*)}   _{ch}$$\sim  $$\rm
M_{Jeans}(n_c)$  at the  typical density  of Galactic  CRDR  dense gas
cores    ($\rm     n_c$$\sim    $(few)$\times    $10$^{5}$\,cm$^{-3}$)
Fig.~\ref{fig:5}    also    shows     that    variations    of    $\rm
U_{CR}/U_{CR,Gal}$$\sim $0.2-10  would leave $\rm  M^{(*)} _{ch}$$\sim
$(0.3--0.5)\,M$_{\odot}$ and still  compatible with a standard stellar
IMF.  This is  important since  a larger  sensitivity of  $\rm M^{(*)}
_{ch}$ on $\rm U_{CR}$ would  have been incompatible with its observed
near-invariance   in   ordinary   star-forming   environments   (where
variations  of the star-formation  rate density  $\dot\rho_{sfr}$, and
thus   of  $\rm   U_{CR}$,  by   factors  of   $\sim  $5-10   are  not
extraordinary).

It must  be noted  that the  choice of the  particular $\rm  (n, T_k)$
phase whose  $\rm M_{Jeans}$ sets  $\rm M^{(*)} _{ch}$ is  not without
ambiguities  even  if  some  concensus  has been  established  on  the
decisive role of $\rm M_{Jeans}({\bf  r},t)$ in driving the outcome of
the  gravoturbulent process at  each spatial  and temporal  point $\rm
({\bf  r},t)$.   Arguments  have  been  made  both  for  $\rm  M^{(*)}
_{ch}$$\propto $$\rm \langle M_{Jeans}\rangle $ at the \emph{onset} of
a gravoturbulently driven  MC fragmentation process (\cite{Kl04}), and
for  $\rm  M^{(*)}   _{ch}$$\sim  $$\rm  M_{Jeans}(n_c,T_k)$  at  some
characteristic  gas density  $\rm  n_c$. For  gravoturbulent-regulated
star-formation  $\rm  n_c$ is  the  density  where efficient  gas-dust
thermal  coupling   of  (far-UV)/optically-shielded  gas   lowers  its
temperature to  the minimum value  possible, i.e. that of  $\sim $$\rm
T_{dust}$,  rendering such  gas  regions nearly  isothermal.  For  the
Galaxy this  occurs at  $\rm n_c$$\sim $10$^{5}$\,cm$^{-3}$,  which is
also  similar to  the density  of  regions where  the transition  from
supersonically   turbulent  to   near  thermal   gas   motions  occurs
(\cite{BerTaf07,Lars95}).

For  CRDRs with  high  average  $\rm U_{CR}$  like  those expected  in
compact  starbursts,  Fig.~\ref{fig:4}  makes  obvious  that  gas-dust
coupling is no  longer adequate to lower $\rm T_{kin}$  to that of the
concomitant dust for densities  as high as $10^{6}$\,cm$^{-3}$. Thus a
characteristic gas  density can no longer  be defined by  the onset of
thermal equilibrium  between gas and  dust in CRDRs, at  least within
the range of densities where  appreciable amounts of mass can be found
in ordinary  MCs. This leaves  the choice of $\rm  M^{(*)} _{ch}$=$\rm
M_{Jeans}(n_c,T_k)$ open, and thus  it is worth examining the possible
direction its value would take  in extreme CRDRs.  In the CR-innudated
H$_2$ clouds inside  compact starbursts $\rm n_c$ can  now be set only
as  the typical  density of  regions where  supersonic  turbulence has
fully  dissipated  and  near-thermal  motions dominate.   The  average
density  of  such  transition  regions  can  be  determined  from  two
well-known  scaling  relations found  for  H$_2$  clouds: $\rm  \sigma
_V(r)$=$\rm  \sigma  _{\circ}  (r/pc)^{h}$ (linewidth-size)  and  $\rm
\langle n  \rangle $=$\rm n_{\circ}  (r/pc)^{-1}$ (density-size) after
setting    $\rm    \sigma    _V$(min)=$\rm   \left(3    k_B    T_k/\mu
m_{H_2}\right)^{1/2}$ as  the minimum linewidth  possible, and solving
for the corresponding mean density,

\begin{equation}
\rm \langle n_{c} \rangle  = n_{\circ} \left(\frac{\mu m_{H_2} \sigma
^{2}       _{\circ}}{3       k_B       T_k}\right)^{1/2h}\sim       11
n_{\circ}\left(\frac{\sigma              _{\circ}}{km\,s^{-1}}\right)^2
\left(\frac{T_k}{10K}\right)^{-1}.
\end{equation}

\noindent
For   h=1/2  (expected   for  pressurized   H$_2$   virialized  clouds
\cite{Elm89}),    $\sigma    _{\circ}$=1.2\,km\,s$^{-1}$   and    $\rm
n_{\circ}$=(few)$\times $10$^3$\,cm$^{-3}$  (from observations), it is
$\rm \langle n_{c} \rangle$$\sim $(few)$\times $10$^4$\,cm$^{-3}$. For
the CR-boosted  $\rm T_{kin}$(min) of extreme  CRDRs $\rm T_k$(min)$\sim
$(50--100)\,K:   $\rm    \langle   n_{c}\rangle   $$\sim   $few$\times
10^3$\,cm$^{-3}$.  Thus the (turbulent gas)$\rightarrow$(thermal core)
transition in extreme CRDRs will occur at lower densities {\it as well
  as}  higher  (CR-boosted)  $\rm  T_{kin}$ values  than  in  ordinary
Galactic ones.  This will then  shift a $\rm M^{(*)} _{ch}$$\sim $$\rm
M_{Jeans}(n_c,T_k)$ both vertically (higher $\rm T_{kin}$), as well as
to the  left (lower $\rm  n(H_2)$) of Fig.~\ref{fig:5},  yielding even
higher $\rm  M^{(*)} _{ch}$  values than those  resulting from  only a
vertical  shift   of  $\rm   M_{Jeans}$  because  of   CR-boosted  gas
temperatures.   Thus  irrespective of  the  H$_2$ cloud  fragmentation
details, and  as long as the Jeans  mass plays a decisive  role in it,
{\it a  much larger characteristic  mass $ M^{(*)} _{ch}$  is expected
  for stars forming in the CRDRs of extreme starbursts.}

\subsection{Ambipolar-diffusion  regulated star formation }

The CR-induced boost of the  ionization fraction x(e) inside the CRDRs
of extreme  starbursts dramatically alters  the star-formation initial
conditions  relevant for the  ambipolar-diffusion (AD)  regulated star
formation scenario. Indeed, the  much larger ionization fractions that
can now be achieved deep inside H$_2$ clouds can in principle: a) keep
the magnetic  field lines strongly ``threaded'' into  molecular gas at
much higher  densities, and  b) as  a result render  much of  its mass
incapable  of star-formation,  at least  in the  simple (AD)-regulated
star  formation  scenario.   This  stems  from  the  now  much  longer
ambipolar      diffusion     timescales:      $\rm     \tau_{AD}$$\sim
$1.6$\times  $$10^{14}$x(e)\,yrs  needed for  a  CR-ionized core  with
density   $\rm   n(H_2)$   to   lose  magnetic   flux   and   collapse
(\cite{McKee89}). Indeed, after inserting x(e) (from Eq. 1.6) it is:

\begin{eqnarray}
\rm \nonumber \tau _{AD}&=& \rm 3.2\times 10^7 r_{gd} \left(\frac{n_{ch}}{2n(H_2)}\right)^{1/2}\\
&\times& \rm \left[\left(1+\frac{n_{ch}}{8n(H_2)}\right)^{1/2}+\left(\frac{n_{ch}}{8n(H_2)}\right)^{1/2}\right] yrs.
\end{eqnarray}

\begin{figure}
\sidecaption
\includegraphics[scale=0.50]{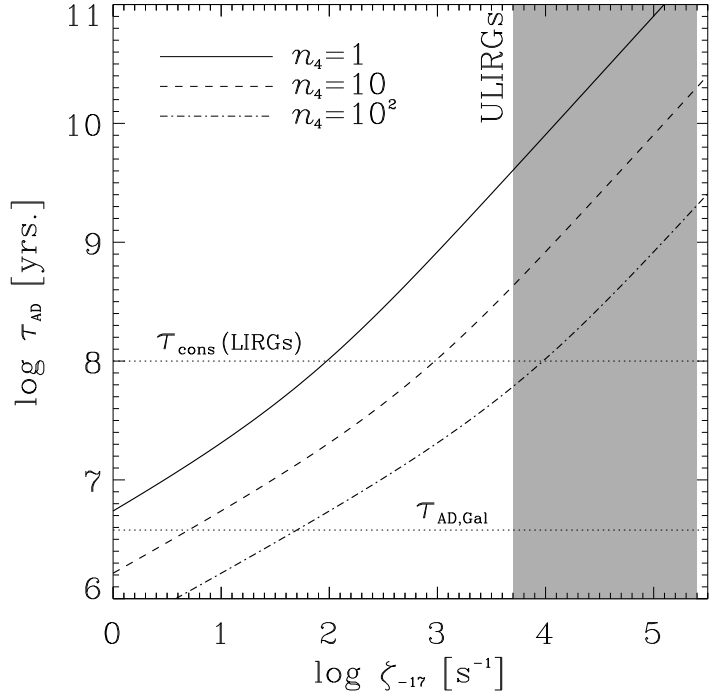}
\caption{The ambipolar  diffusion (AD) timescale as a  function of the
  CR     ionization      rate     $\rm     \zeta_{-17}$=$\rm     \zeta
  _{CR}/(10^{-17}s^{-1})$   (assumed  $\propto  $$\rm   U_{CR}$),  for
  n=[10$^{4}$,   10$^{5}$,    10$^{6}$]\,cm$^{-3}$   ($\rm   n_4$=$\rm
  n(H_2)/(10^4\,cm^{-3})$).  For  comparison the AD  timescale for ISM
  conditions typical in the Galaxy  is also shown as a horizontal line
  at  the  bottom.  The  H$_2$  gas  consumption  timescale $\rm  \tau
  _{cons}$$\sim $M(H$_2$)/SFR for a  typical (U)LIRG is also marked by
  the higher horizontal line.  The  shaded area marks the region where
  such galaxies are  expected to be in terms of  the $\rm \zeta _{CR}$
  in   their  CRDRs,  where   clearly  $\rm   \tau_{AD}$$>$$\rm  \tau
  _{cons}$. }
\label{fig:6}       
\end{figure}

\noindent
In Fig.~\ref{fig:6} $\rm \tau _{AD}$ is  shown as a function of the CR
ionization rate $\rm \zeta_{CR}$  (assumed $\propto $$\rm U_{CR}$) for
three typical  densities of cloud  cores inside CRDRs.  It  is obvious
that the much higher $\rm  \zeta_{CR}$ values expected in the CRDRs of
(U)LIRGs will make the  corresponding $\rm \tau _{AD}$ timescales much
longer than those  expected in the Galaxy.  Moreover  $\rm \tau _{AD}$
in  (U)LIRGs  will be  longer  than  their  so-called gas  consumption
timescales   $\rm   \tau    _{cons}$=$\rm   M(H_2)$/SFR,   which   are
observationally determined using $\rm M(H_2)$ (usually obtained via CO
J=1--0) and the star  formation rate (SFR) (usually obtained from~$\rm
L_{IR}$).

For  any theory  of star  formation  $\rm \tau  _{cons}$ represents  a
natural  \emph{maximum} for  any theoretically  deduced characteristic
star-formation   timescale  $\rm  \tau_{*,ch}$   since  star-formation
feedback (via SNR-induced shocks, O,B star far-UV radiation) will only
lenghten  any such $\rm  \tau_{*,ch}$ by  destroying the  H$_2$ clouds
fueling star  formation (which  then have to  reform to  be SF-capable
again).  From Fig.~\ref{fig:6} it is obvious that for (U)LIRGs it will
always  be  $\rm   \tau_{AD}$$>$$\rm  \tau  _{cons}$,  thus  rendering
AD-regulated  star  formation as  unrealistic.   For  the average  ISM
conditions in such vigorously  star-forming galaxies a process such as
reconnection diffusion (\cite{Laz11}) can still provide magnetic field
diffusion  fast  enough  as  to  keep  $\rm  \tau_{*,ch}$$<$$\rm  \tau
_{cons}$  (\cite{Laz12}).

 In summary, in the AD-diffusion regulated star-formation scenario, as
 in the gravoturbulent one, \emph{the large CR energy densities in the
   CRDRs  of compact  starbursts will  dramatically alter  the initial
   conditions  for  star  formation.}   The effects  on  the  emergent
 stellar   IMF  are   however   less  clear   to   discern  than   for
 gravoturbulently-regulated   star  formation,   especially   if  fast
 reconnection  diffusion  plays  a  role in  removing  magnetic  field
 support  from  dense  H$_2$  gas  with  orders  of  magnitude  larger
 ionization fractions than in ordinary star-forming galaxies.

\subsection{CR-induced effects on the  equation of state of the H$_2$ gas}

Another  effect of  the  large CR  energy  densities in  the CRDRs  of
compact starbursts  is \emph{the  erasing of the  so-called inflection
  point of  the effective equation of  state (EOS) of  the H$_2$ gas.}
The latter,  parametrized as a polytrope P=K$\rho  ^{\gamma}$, is used
in  many  numerical  simulations  of self-gravitating  turbulent  gas.
These found its polytropic index  $\gamma $, and more specifically the
so-called EOS  inflection point (i.e. the  characteristic density $\rm
n_c$ where $\gamma $$<1$ for n$<$$\rm n_c$  flips to $\gamma $$\geq
$1 for n$\geq  $$\rm n_c$), playing a crucial  role in determining the
functional shape and mass scale  of the resulting mass spectrum of the
collapsed  dense gas  cores (and  thus the  stellar  IMF) (\cite{Li03,
  Japp05}).  For ordinary ISM  conditions this inflection point occurs
within  CRDRs at  $\rm  n_c$$\sim $10$^{5}$\,cm$^{-3}$  (\cite{Lars95,
  Japp05}) because of the onset of strong thermal gas-dust coupling in
the  (far-UV)/optically-shielded gas regions,  which makes  them nearly
isothermal and sets  a characteristic IMF mass scale  of $\rm M^{(ch)}
_{*}$$\sim $$\rm M_{jeans}(n_c)$.

\begin{figure}
\sidecaption
\includegraphics[scale=0.50, angle=90]{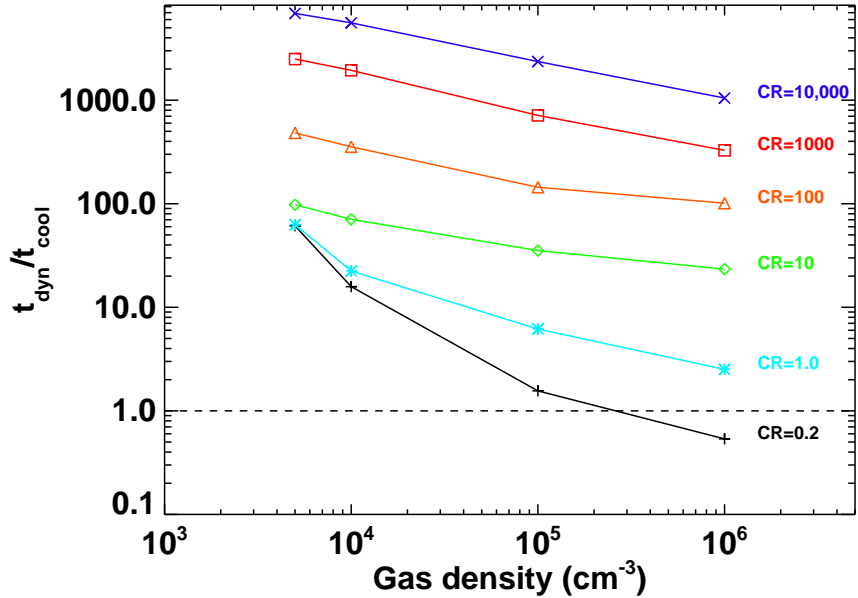}
\caption{The ratio  of dynamical/cooling  timescales for H$_2$  gas in
  CRDRs, for CR  energy densities expected in star-formation-quiescent
  up to starburst galaxies (scaled to the average Galactic~value).}
\label{fig:7}       
\end{figure}

In the CRDRs  of compact starbursts the H$_2$ gas  and the dust remain
thermally decoupled with  $\rm T_{kin}$$>$$\rm T_{dust}$ for densities
as high as n$\sim $10$^{6}$\,cm$^{-3}$ (Fig.~\ref{fig:4}), erasing the
EOS  inflection  point.   This  maintains  strong  H$_2$  gas  cooling
($\propto  $n$^{2}$)  at much  higher  densities  ($\geq $$\rm  n_c$),
rendering local  cooling timescales ($\rm \tau  _{cool}$) much shorter
than   the  dynamical   ones   ($\rm  \tau   _{dyn}$).   This   allows
thermodynamics  to  always  stay  a  step  ahead  of  self-gravity  in
decreasing the  local Jeans mass (as densities  climb and temperatures
drop) faster  than the  gravitationally-driven evolution of  the cloud
fragments it  to mass fragments of $\sim  $$\rm M_{jeans}({\bf r},t)$.
A comparison of these two important timescales as a function of the CR
energy densities  is shown in Fig.~\ref{fig:7}, from  where it becomes
obvious   that  an   inflection  point   where   $\rm  \tau_{dyn}/\tau
_{cool}$$\sim    $1    is    not    even    approached    once    $\rm
U_{CR}/U_{CR,Gal}$$\geq  $10,  even for  densities  as  high as  $\sim
$10$^{6}$\,cm$^{-3}$.  Clearly  \emph{any future numerical simulations
  of molecular clouds in extreme CRDRs must take this into account, or
  abandon the use of an EOS for the gas altogether.}

\section{A CR-controlled stellar IMF: some consequences}

\begin{figure}
\sidecaption
\includegraphics[width=0.95\textwidth,trim=55 0 55 0]{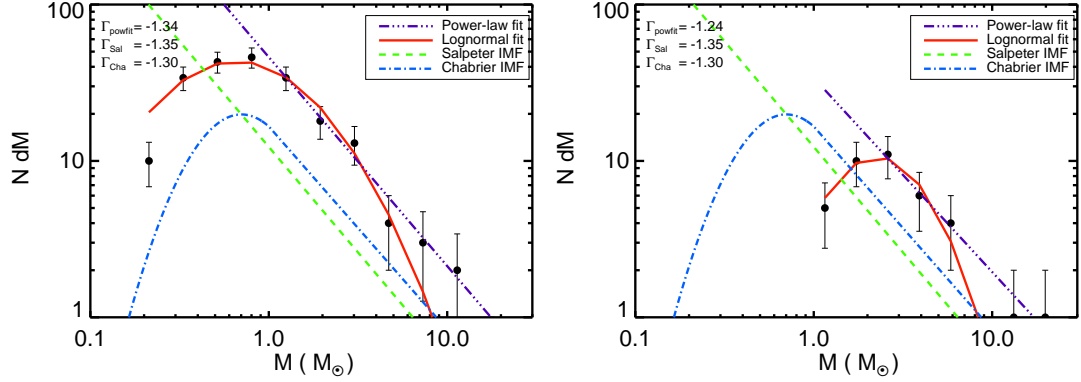}
\caption{The  stellar  initial mass  functions  (IMFs) from  numerical
  simulations  of  turbulent   H$_2$  clouds,  for  $\rm  U_{CR}$=$\rm
  U_{CR,Gal}$    (left),     100$\times$$\rm    U_{CR,Gal}$    (right)
  \cite{Hoc11}.   For  comparison  purposes,  a  Salpeter-type  (green
  dashed) and Chabrier-type (blue dot-dashed) IMF fits to the emergent
  mass spectrum are also shown along with a linear fit and a lognormal
  fit  (purple  and  red  lines).   The  power-law  slopes  above  the
  turn-over  mass $\rm  M^{(*)} _{ch}$  are  given in  the upper  left
  corner.  }
\label{fig:8}       
\end{figure}

A top-heavy stellar IMF can be the result a higher characteristic mass
$\rm M^{(*)}  _{ch}$ for  the young stars  (the so-called  IMF `knee`,
which  is $\sim  $(0.3--0.5)\,$\rm  M_{\odot}$ for  the  Galaxy) or  a
shalower power  law than  its Salpeter value  of $\alpha  $$\sim $2.35
(for an  IMF functional  form of $\rm  dN(M_{*})/dM_{*}$$\propto $$\rm
M_{*}  ^{-\alpha}$  and  stellar  masses  between  1\,M$_{\odot}$  and
100\,M$_{\odot}$).  In  a gravoturbulent scenario the  exact shape and
mass scale of  the emergent stellar IMF under  the very different star
formation initial  conditions in extreme  CRDRs of ULIRGs can  only be
determined with new H$_2$ cloud simulations that make use of these new
conditions.   In the  past simulations  using so-called  starburst ISM
conditions  have been done,  recovering large  $\rm M^{(*)}  _{ch}$ in
such environments (\cite{Kl07}).   Unfortunately such simulations used
conditions typical of  PDRs rather than CRDRs, an  approach that would
have then produced a varying IMF even in the Milky Way given the large
range of  gas properties in  its PDRs. It  was only recently  that the
effects of large CR energy  densities on H$_2$ cloud fragmentation and
the  stellar IMF  have  been explored  (\cite{Hoc11})  and found  $\rm
M^{(*)}     _{ch}$     boosted     to    $\sim     $(2-3)\,M$_{\odot}$
(Fig.~\ref{fig:8}).

The dependance of the thermal and ionization state of the gas in CRDRs
on the average  $\rm U_{CR}$ permeating the ISM  of a galaxy naturally
links the SF  initial conditions and the resulting  IMF to the average
SFR   density    $\dot\rho_{sfr}$,   provided   $\rm   U_{CR}$$\propto
$$\dot\rho_{sfr}$ (with the  details of CR transport/escape mechanisms
in  quiescent  SF disks  and  starbursts  setting the  proportionality
factor).   CR-controlled   star  formation  initial   conditions  then
naturally yield a bimodal stellar IMF in galaxies, with high values of
$\dot\rho_{sfr}$ determining the branching point, namely:\\

\noindent
  (merger-driven star formation)$\rightarrow $(high $\dot\rho_{sfr}$)$\rightarrow$(high
    $\rm  U_{CR}$)$\rightarrow $(top heavy IMF),\\
\noindent
(isolated gas-rich disk star formation)$\rightarrow $(low $ \dot\rho_{sfr}$)$\rightarrow$(low
    $\rm  U_{CR}$)$\rightarrow $(Galactic IMF).\\

\noindent
Such  an IMF  bimodality, with  a top-heavy  IMF  in merger/starbursts
versus a  regular IMF in  isolated gas-rich disks has  been postulated
for  some time  in  hierarchical galaxy  formation  models seeking  to
explain their relative populations  across cosmic time in $\Lambda$CDM
cosmologies  (\cite{Baugh05}).    \emph{CR-controlled  star  formation
  initial conditions in  galaxies can now set this  on a firm physical
  basis for the first time.}  It must be noted that a top-heavy IMF in
high-$\dot\rho_{sfr}$ systems  such as compact  starbursts in mergers,
will also  have serious  implications on the  SFRs deduced  from their
observed IR luminosities  as such IMFs will have  several times higher
energy  outputs per  stellar mass  than  the ordinary  one.  Then  the
tremendeous  SFRs deduced  for merger/starbursts,  especially  at high
redshifts  ($\sim $10$^{3}$\,M$_{\odot}$\,yr$^{-1}$), may  actually be
$\sim $(3-5) times lower.

Finally   an  $\dot\rho_{sfr}$-dependant   IMF  as   the   outcome  of
CR-controlled star formation initial  conditions in galaxies will also
naturally yield  integrated galactic IMFs (IGIMFs) that  depend on the
star  formation history  (SFH) of  galaxies.  This  is  simply because
$\dot\rho_{sfr}$  can  change  significantly during  the  evolutionary
track of a galaxy, especially during its early gas-rich epoch, even in
the absence of  mergers. Such a SFH-dependance of  the IGIMFs has been
considered  as  the   cause  behind  the  well-known  mass-metallicity
relation of galaxies (\cite{Krou07}).

\section{Future outlook}

The unique  signatures imparted on  the thermal and chemical  state of
the  H$_2$ gas  in extreme  CRDRs  have now  been studied  extensively
(\cite{Pap10,Mej11,Bay11}), and most of them will be easily accessible
in the new era of ALMA  whose commissioning is now ongoing at Llano de
Chajnantor on the Atacama  Desert Plateau.  Numerical studies of H$_2$
clouds without  the use  of an EOS  for the  gas, that take  also into
account  the strongly  ´anchored´ magnetic  fields on  dense  gas with
large  ionization fractions  are now  becoming  possible.  High-energy
$\gamma$-ray observations now started  directly probing the average CR
energy densities  of other galaxies  (\cite{Acciari09, Acero09}). Thus
we are looking at a decade where the properties of CRDRs, their effect
on the stellar IMF of galaxies, and their role in galaxy formation can
be  decisively explored  with powerful  theoretical  and observational
tools.   Finally  these  explorations  will  critically  benefit  from
studies of the stellar IMF and its dependence on ambient conditions in
galaxies   that   are  now   converging   to   much  sharper   picture
(\cite{Krou11}).

\begin{acknowledgement}
Padelis P. Papadopoulos  would like to warmly thank  the organizers of
the  conference in  Sant Cugat  for an  exciting week  where  much was
learned  and new vistas  for both  high energy  and the  so-called low
energy Astrophysics  came into view.  This project was also  funded by
the John Latsis Public Benefit Foundation. The sole responsibility for
the content lies with its authors.
\end{acknowledgement}

\newpage

\section*{Appendix}
\addcontentsline{toc}{section}{Appendix}
The CR heating rate used in this work is given by

\begin{equation}
\rm \Gamma_{\mathrm{CR}}=1.95\times10^{-28}n_{\mathrm{H}}
 \left(\frac{\zeta_{CR}}{1.3\times10^{-17}\,s^{-1}}\right)\, ergs\, cm^{-3}\,s^{-1}
\end{equation}

\noindent
 where $\rm  \zeta _{CR}$$\propto $$\rm  U_{CR}$ being the  Cosmic Ray
 ionization rate (in s$^{-1}$), with an adopted Galactic value of $\rm
 \zeta _{CR,Gal}$=5$\times$10$^{-17}$  s$^{-1}$ (corresponding to $\rm
 U_{CR}$=$\rm  U_{CR,Gal}$),  and   $\rm  n_{H}$=2n(H$_2$)  for  fully
 molecular  gas.  For  an optically  thin  OI line  (and thus  maximal
 cooling)        $\rm        \Lambda_{\mathrm{OI63}}=\chi_O        n_H
 C_{\mathrm{lu}}E_{\mathrm{ul}}$, which becomes

\begin{equation}
\rm \Lambda_{\mathrm{OI63}}= 3.14\times10^{-14}\chi_O
\rm n_{\mathrm{H}}C_{\mathrm{ul}}\left(\frac{g_{\mathrm{u}}}{g_{\mathrm{l}}}\right)
\rm \exp{\left(-227.72/T_{\mathrm{kin}}\right)}\,ergs\,
\end{equation}

\noindent
where      $g_{\mathrm{u}}$=3     and      $g_{\mathrm{l}}$=5,     and
$\chi_{\mathrm{O}}$=[O/H]  being the  abundance of  oxygen  not locked
onto    CO    ($\chi_{\mathrm{O}}$$\sim$4.89$\times$10$^{-4}$).    The
collisional de-excitation coefficient is given by (\cite{Lis99})

\begin{equation}
\rm C_{\mathrm{ul}}=n_{\mathrm{H}} 10^{0.32\log{T_{\mathrm{kin}}}-10.52}=
\rm     3.02\times10^{-11}n_{\mathrm{H}}T_{\mathrm{kin}}^{0.32}\,cm^{-3}\,s^{-1},
\end{equation}

\noindent
 Thus finally Equation 1.11 becomes,

\begin{equation}
\rm \Lambda_{\mathrm{OI63}}=2.78\times10^{-28}n_{\mathrm{H}}^2T_{\mathrm{kin}}^{0.32}
\rm \exp{\left(-227.72/T_{\mathrm{kin}}\right)}\,ergs\,cm^{-3}\,s^{-1}.
\end{equation}

\noindent
  For densities  of $n_{\mathrm{H}}<$10$^{5}$ cm$^{-3}$  and strong CR
  fluxes,  a small  fraction of  carbon remains  in the  form  of C$^{+}$,
  acting as a coolant with

\begin{equation}
\rm \Lambda_{\mathrm{C^{+}}}=1.975\times10^{-23}n_{\mathrm{H}}^2\chi_{\mathrm{C^{+}}}
\rm \exp{\left(-92.2/T_{\mathrm{kin}}\right)}\,ergs\,cm^{-3}\,s^{-1},
\end{equation}

\noindent
computed in a similar fashion as the OI(63\,$\mu $m) line cooling, and
for a  fully transparent medium.  Gas-grain  accommodation cooling the
gas  depending on  their temperature  difference can  be  expressed as
(\cite{Bur83}):

\begin{equation}
\rm \Lambda_{\mathrm{g-d}}=4.0\times10^{-12}\alpha n_{\mathrm{H}}
\rm n_{\mathrm{d}}\sigma_{\mathrm{g}}\sqrt{T_{\mathrm{kin}}}
\rm \left(T_{\mathrm{kin}}-T_{\mathrm{dust}}\right)
\end{equation}

\noindent
 where  $n_{\mathrm{g}}$ is  the  number density  of  dust grains  and
$\sigma_{\mathrm{g}}$ is  the grain cross-section  (cm$^{-2}$). If the
gas-to-dust    ratio   is    100,   the    dust   mass    density   is
$\rho_{\mathrm{d}}$=3.5\,g\,cm$^{-3}$,   and    a   dust   radius   of
$a=$~0.1~$\mu$m then

\begin{equation}
\rm n_{\mathrm{d}}=2.2\times 10^{-2}\times\mu_{\mathrm{H}}
\rm n_{\mathrm{H}}/(4/3\pi\rho_{\mathrm{d}}a^{3})=7.88\times10^{-12}n_{\mathrm{H}}.
\end{equation}

\noindent
 The gas-grain accommodation factor $\alpha $ is given by (\cite{Gro94}):

\begin{equation}
\rm \alpha=0.35\exp{\left(-\sqrt{\frac{T_{\mathrm{dust}}+T_{\mathrm{kin}}}{500}}\right)},
\end{equation}

\noindent
which we  set to  the maximum value  $\alpha$=0.35 (i.e.   maximum gas
cooling from gas-dust interaction).  Thus  the cooling term due to the
gas-dust interaction (Equation 1.15) becomes

 \begin{equation}
\rm \Lambda_{\mathrm{g-d}}=3.47\times10^{-33}n_{\mathrm{H}}^{2}
\rm \sqrt{T_{\mathrm{kin}}}\left(T_{\mathrm{kin}}-T_{\mathrm{dust}}\right) ergs\,cm^{-3}\,s^{-1}.
\end{equation}

\noindent
Finally detailed PDR code models allow us to parametrize  the cooling 
due to the CO rotational transitions as

\begin{equation}
\rm \Lambda_{\mathrm{CO}}=4.4\times10^{-24}\left(\frac{n_{\mathrm{H}}}{10^4\,cm^{-3}}\right)^{3/2}
\rm \left(\frac{T_{\mathrm{kin}}}{10\,K}\right)^2
\rm \left(\frac{\chi_{\mathrm{CO}}}{\chi_{\mathrm{[C]}}}\right)\,ergs\,cm^{-3}\,s^{-1},
\end{equation}

\noindent
where                              we                              set
$\chi_{\mathrm{CO}}/\chi_{\mathrm{[C]}}$=(0.97,0.98,0.99,1.0)       for
densities of 5$\times$10$^{3}$, 10$^{4}$, 10$^{5}$, 10$^{6}$ cm$^{-3}$
respectively      (with     $\chi_{\mathrm{CO}}/\chi_{\mathrm{[C]}}$=1
corresponding to all carbon locked onto CO).


\vspace*{1cm}

\begin{thebibliography}{}%
%
%
\bibitem{HolTiel99} Hollenbach D. J., \& Tielens A. G. G. M.
Reviews of Modern Physics {\bf Vol. 17}, No1, January 1999\\
\bibitem{Wolf03}  Wolfire M. G., McKee C. F.,  Hollenbach D., 
 Tielens, A. G. G. M. 2003, Astrophysical Journal,  {\bf 587}, 278\\
\bibitem{TiHol95a} Tielens A. G. G. M. \& Hollenbach D. J. 1985a
Astrophysical Journal, {\bf 291}, 722\\
\bibitem{TiHol95b} Tielens A. G. G. M. \& Hollenbach D. J. 1985b
Astrophysical Journal, {\bf 291}, 747\\
\bibitem{GaoSol04} Gao Yu, \& Solomon P. M. 2004, Astrophysical
Journal Supplement, {\bf  152}, 63\\
\bibitem{Sol92} Solomon P. M.,  Downes D., \& Radford S. J. E 1992,
Astrophysical Journal Letters {\bf 387}, 55\\
\bibitem{PaNo02} Padoan P., \& Nordlund A. 2002, Astrophysical Journal, {\bf 576}, 870\\
\bibitem{PanPad09} Pan L., \& Padoan P. 2009, Astrophysical Journal, {\bf 692}, 594\\
\bibitem{Pel06} Pelupessy F. I., Papadopoulos P. P., \&  van der Werf P. P. 2006,
Astrophsyical Journal, {\bf 645}, 1024\\
\bibitem{BerTaf07} Bergin, E. A. \& Tafalla, M. 2007, Annual Reviews of Astronomy
\& Astrophysics, {\bf 45}, 339\\ 
\bibitem{Pin07} Pineda, J. L. \& Bensch, F. 2007, Astronomy \& Astrophysics, {\bf 470}, 615\\
\bibitem{Pin10} Pineda, J. E., Goodman, A. A, Arce, H. E. et al. 2010, Astrophysical
 Journal Letters, {\bf 712}, L116\\ 
\bibitem{Gold78} Goldsmith, P. F., \&  Langer, W. D. 1978, Astrophysical Journal, {\bf 222}, 881\\
\bibitem{Oss02} Ossenkopf, V. 2002, Astronomy \& Astrophysics, {\bf 391}. 295\\
\bibitem{Gold01} Goldsmith, P. F. 2001, Astrophysical Journal, {\bf 557}, 736\\
\bibitem{Lis99} Liseau, R. et al. 1999, Astronomy \& Astrophysics 1999, {\bf 344}, 342\\
\bibitem{Bur83} Burke, J. R., \& Hollenbach, D. J. 1983, Astrophysical Journal, {\bf 265}, 223\\
\bibitem{Gro94} Groenewegen, M. A. T. 1994, Astronomy \& Astrophysics, {\bf 290}, 531\\
\bibitem{Pap11} Papadopoulos, P. P., Thi. W.-F., Miniati, F., \& Viti S. 2011, Monthly
Notices of the Royal Astronomical Society, {\bf 414}, 1705\\
\bibitem{McKee89} McKee, C. F. 1989, Astrophysical Journal, 1989, Astrophysical
Journal, {\bf 345}, 342\\
\bibitem{Mous76} Mouschovias, T. Ch., \&  Spitzer L. Jr, 1976, Astrophsyical
Journal, {\bf 210}, 326\\
\bibitem{Lars95} Larson, R. B. 2005, Monthly Notices of the Royal Astronomical Society,
{\bf 359}, 211\\
\bibitem{Elm08} Elmegreen B. G., Klessen R. S., \& Wilson C. D. 2008, Astrophysical Journal,
 {\bf 681}, 365\\
\bibitem{Sak08} Sakamoto, K. et al. 2008, Astrophysical Journal, {\bf 684}, 957\\
\bibitem{DS98} Downes D., \& Solomon P. M. 1998, Astrophysical Journal, {\bf 507}, 615\\
\bibitem{Gre09} Greve T. R., Papadopoulos P. P., Gao Y., \& Radford S. J. E. 2009,
 Astrophysical Journal, {\bf 692}, 1432\\
\bibitem{Pap12} Papadopoulos P. P., van der Werf P. P., Xilouris E. M., Isaak K. G.,
Gao Y., \& Muehle S. 2012, Monthly Notices of the Royal Astronomical Society, (in press, arXiv:1109.4176)\\
\bibitem{Rang12} Rangwala N. et al. 2012, Astrophysical Journal, {\bf 743}, 94\\
\bibitem{Pap10} Papadopoulos P. P., 2010, Astrophysical Journal, {\bf 720}, 226\\
\bibitem{Ys07}  Yusef-Zadeh F., Wardle M., Roy S. 2007, Astrophysical Journal Letters {\bf 665,} L123\\
\bibitem{Brad03}  Bradford C. M., Nikola T., \& Stacey G. J. 2003, Astrophysical Journal {\bf 586}, 891\\
\bibitem{Hail08} Hailey-Dunsheath S., Nikola T., Stacey G. J. et al. 2008, Astrophysical Journal Letters, 
{\bf 689,} L109\\
\bibitem{Kl04} Klessen R. S. 2004, Astrophysics and Space Science, {\bf 292}, Issue 1, p. 215\\ 
\bibitem{Elm08} Elmegreen B. G., Klessen R. S., \& Wilson C. D. 2008, Astrophysical
 Journal, {\bf 681}, 365\\
\bibitem{Elm89} Elmegreen B. G. 1989, Astrophysical Journal, {\bf 338}, 178\\
\bibitem{Laz11} Lazarian A. 2011, Nonlin. Processes Geophys. (in press, arXiv:1111.0694)\\
\bibitem{Laz12} Lazarian A.\, Esquivel A., \& Crutcher R. 2012, Astrophysical Journal (arXiv:1206.4698)\\
\bibitem{Li03} Li Y., Klessen R. S., \& McLow M.-M. 2003, Astrophysical Journal, {\bf 592}, 975\\
\bibitem{Japp05} Jappsen A.-K., Klessen R. S., Larson R. B., Li Y., McLow M.-M. 2005, 
                 Astronomy \& Astrophysics, {\bf 435} 611\\
\bibitem{Kl07}	Klessen R. S.,  Spaans M. \& Jappsen A.-K. 2007, Montly Notices of the
                Royal Astronomical Society, {\bf 374}, L29\\
\bibitem{Hoc11} Hocuk S., \& Spaans M. 2011, Astronomy \& Astrophysics, {\bf 536}, 41\\
\bibitem{Baugh05} Baugh C. M. et al. 2005, Monthly Notices of the Royal Astronomical
Society, {\bf 356}, 1191\\
\bibitem{Krou07} Koeppen J., Weidner C., \& Kroupa P. 2007, Monthly Notices of the Royal
Astronomical Society, {\bf 375}, 673\\ 
\bibitem{Mej11} Meijerink R., Spaans M., Loenen A. F., \& van der Werf P. P. 2011,
Astronomy \& Astrophysics, {\bf 525}, 119\\
\bibitem{Bay11} Bayet E., Williams D. A., Hartquist T. W., \& Viti S. 2011,
Monthly Notices of the Royal Astronomical Society, {\bf 414}, 1583\\ 
\bibitem{Acciari09} Acciari V. A. et al. 2009,  Nature, {\bf 462}, 770\\
\bibitem{Acero09} Acero F. et al. 2009, Science, {\bf 326}, 1080\\
\bibitem{Krou11} Kroupa P. et al. 2012, in {\it Stellar Systems and
Galactic Structure}, Vol. V. (in press, arXiv:1112.3340)\\
\end{thebibliography}
\end{document}